\documentclass{PoS}

\title{Dynamical lattice computation of the Isgur-Wise functions $\tau_{1/2}$ and $\tau_{3/2}$}

\ShortTitle{Dynamical lattice computation of the Isgur-Wise functions $\tau_{1/2}$ and $\tau_{3/2}$}

\author{Benoit Blossier\\
        Laboratoire de Physique Th\'eorique (B\^at.210), Universit\'e Paris-Sud XI, Centre d'Orsay, 91405 Orsay-Cedex, France\\
        E-mail: \email{benoit.blossier@th.u-psud.fr}}

\author{\speaker{Marc Wagner}\\
        Humboldt-Universit\"at zu Berlin, Institut f\"ur Physik, Newtonstra{\ss}e 15, D-12489 Berlin, Germany\\
        E-mail: \email{mcwagner@physik.hu-berlin.de}}

\author{Olivier P\`ene\\
        Laboratoire de Physique Th\'eorique (B\^at.210), Universit\'e Paris-Sud XI, Centre d'Orsay, 91405 Orsay-Cedex, France\\
        E-mail: \email{olivier.pene@th.u-psud.fr}}

\abstract{We perform a two-flavor dynamical lattice computation of the Isgur-Wise functions $\tau_{1/2}$ and $\tau_{3/2}$ at zero recoil in the static limit. We find $\tau_{1/2}(1) = 0.297(26)$ and $\tau_{3/2}(1) = 0.528(23)$ fulfilling Uraltsev's sum rule by around 80\%. We also comment on a persistent conflict between theory and experiment regarding semileptonic decays of $B$ mesons into orbitally excited $P$ wave $D$ mesons, the so-called ``$1/2$ versus $3/2$ puzzle'', and we discuss the relevance of lattice results in this context.}

\FullConference{The XXVII International Symposium on Lattice Field Theory\\
		 July 26-31, 2009\\
		 Peking University, Beijing, China}

\begin{document}


\section{Introduction}

We are concerned with semileptonic decays of $B$ mesons ($B$ and $B^\ast$) into orbitally excited $P$ wave $D$ mesons (collectively denoted as $D^{\ast \ast}$'s): $B^{(\ast)} \rightarrow D^{\ast \ast} \, l \, \nu$. These decays are of particular interest, because there is a persistent conflict between theory and experiment, the so-called ``$1/2$ versus $3/2$ puzzle'': while experimental results indicate that a decay into ``$1/2$ $P$ wave $D^{\ast \ast}$'s'' is more likely, theory favors the decay into ``$3/2$ $P$ wave $D^{\ast \ast}$'s'' (for recent reviews cf.\ \cite{Uraltsev:2004ra,Bigi:2007qp}).


\subsection{Heavy-light mesons}

A heavy-light meson is made from a heavy quark ($b$, $c$) and a light quark ($u$, $d$), i.e.\ \\ $B = \{\bar{b} u \, , \, \bar{b} d\}$ and $D = \{\bar{c} u \, , \, \bar{c} d\}$.

In the static limit ($m_b , m_c \rightarrow \infty$) there are no interactions involving the static quark spin. Therefore, it is appropriate to classify states according to parity $\mathcal{P}$ and the total angular momentum of the light quarks and gluons $j$ (cf.\ the left column of Table~\ref{TAB001}).

If $m_b , m_c$ are finite, $j$ is not a good quantum number anymore. States have to be classified according to parity $\mathcal{P}$ and total angular momentum $J$ (cf.\ the right column of Table~\ref{TAB001}). Although $j$ is not a ``true quantum number'' anymore, it is still an approximate quantum number justifying the notation $D_J^j$. The above mentioned $P$ wave $D^{\ast \ast}$'s are $\{ D_0^\ast \, , \, D'_1 \, , \, D_1 \, , \, D_2^\ast \} = \{ D_0^{1/2} \, , \, D_1^{1/2} \, , \, D_1^{3/2} \, , \, D_2^{3/2} \}$.

\begin{table}[htb]
\begin{center}
\begin{tabular}{|l|l|}
\hline
 & \vspace{-0.50cm} \\
$j^\mathcal{P}$ & $J^\mathcal{P}$ \\
 & \vspace{-0.50cm} \\
\hline
 & \vspace{-0.50cm} \\
$(1/2)^- \ \equiv \ S$ & $0^- \ \equiv \ B, D$ \\
 & $1^- \ \equiv \ B^\ast, D^\ast$ \\
 & \vspace{-0.50cm} \\
\hline
 & \vspace{-0.50cm} \\
$(1/2)^+ \ \equiv \ P_-$ & $0^+ \ \equiv \ D_0^\ast \ \equiv \ D_0^{1/2}$ \\
 & $1^+ \ \equiv \ D'_1 \ \equiv \ D_1^{1/2}$ \\
$(3/2)^+ \ \equiv \ P_+$ & $1^+ \ \equiv \ D_1 \ \equiv \ D_1^{3/2}$ \\
 & $2^+ \ \equiv \ D_2^\ast \ \equiv \ D_2^{3/2}$\vspace{-0.50cm} \\
 & \\
\hline
\end{tabular}
\end{center}

\caption{\label{TAB001}Classification of heavy-light mesons (left: static limit; right: finite heavy quark masses).}
\end{table}


\subsection{The $1/2$ versus $3/2$ puzzle}

Experiments (ALEPH, BaBar, BELLE, CDF, DELPHI, D{\O}), which have studied the semileptonic decay $B \rightarrow X_c \, l \, \nu$ (where $X_c$ is some hadronic part containing a $c$ quark), find the following composition of $X_c$:
\begin{itemize}
\item $\approx 75\%$ $D$ and $D^\ast$, i.e.\ $S$ wave states (which is in agreement with theory).

\item $\approx 10\%$ $D_1^{3/2}$ and $D_2^{3/2}$, i.e.\ $j = 3/2$ $P$ wave states (which is in agreement with theory).

\item For the remaining $\approx 15\%$ the situation is rather vague: a natural candidate would be $D_0^{1/2}$ and $D_1^{1/2}$, i.e.\ $j = 1/2$ $P$ wave states. This, however, would imply \\ $\Gamma(B \rightarrow D_{0,1}^{1/2} \, l \, \nu) > \Gamma(B \rightarrow D_{1,2}^{3/2} \, l \, \nu)$, which is in conflict with theory. This conflict between experiment and theory is called the $1/2$ versus $3/2$ puzzle.
\end{itemize}

On the theory side most statements are made in the static limit $m_b,m_c \rightarrow \infty$. In this limit the eight matrix elements relevant for decays $B \rightarrow D^{\ast \ast} \, l \, \nu$ can be parameterized by two form factors, the Isgur-Wise functions $\tau_{1/2}$ and $\tau_{3/2}$ \cite{Isgur:1990jf}. Here we only list two of these matrix elements:
\begin{eqnarray}
\label{EQN002} & & \hspace{-0.7cm} \langle D_0^{1/2}(v') | \bar{c} \gamma_5 \gamma_\mu b | B(v) \rangle \ \ \propto \ \ \tau_{1/2}(w) (v - v')_\mu \\
\label{EQN003} & & \hspace{-0.7cm} \langle D_2^{3/2}(v',\epsilon) | \bar{c} \gamma_5 \gamma_\mu b | B(v) \rangle \ \ \propto \ \ \tau_{3/2}(w) \Big((w+1) \epsilon^\ast_{\mu \alpha} v^\alpha - \epsilon^\ast_{\alpha \beta} v^\alpha v^\beta v'_\nu\Big) ,
\end{eqnarray}
where $v$ and $v'$ are the four velocities associated with the $B$ and the $D$ meson respectively, \\ $w = (v' \cdot v)$ and $\epsilon$ is the polarization tensor of the $D$ meson.

By means of operator product expansion (OPE) a couple of sum rules has been derived in the static limit \cite{LeYaouanc:1996bd,Uraltsev:2000ce}. The most prominent in this context is the Uraltsev sum rule,
\begin{eqnarray}
\label{EQN001} \sum_n \left(\Big|\tau_{3/2}^{(n)}(1)\Big|^2 - \Big|\tau_{1/2}^{(n)}(1)\Big|^2\right) \ \ = \ \ \frac{1}{4} ,
\end{eqnarray}
where $\tau_{1/2} \equiv \tau_{1/2}^{(0)}$, $\tau_{3/2} \equiv \tau_{3/2}^{(0)}$ and the sum is over all $1/2$ and $3/2$ $P$ wave states respectively. From experience with sum rules one expects approximate saturation from the ground states, i.e.\
\begin{eqnarray}
\Big|\tau_{3/2}^{(0)}(1)\Big|^2 - \Big|\tau_{1/2}^{(0)}(1)\Big|^2 \ \ \approx \ \ \frac{1}{4} ,
\end{eqnarray}
which implies $|\tau_{1/2}(1)| < |\tau_{3/2}(1)|$. This in turn strongly suggests \\ $\Gamma(B \rightarrow D_{0,1}^{1/2} \, l \, \nu) < \Gamma(B \rightarrow D_{1,2}^{3/2} \, l \, \nu)$, which, as already mentioned, is in conflict with experiment.

Phenomenological models \cite{Morenas:1997nk,Ebert:1998km} give the same qualitative picture, even when considering finite heavy quark masses \cite{Ebert:1999ga}.

Possible explanations to resolve the $1/2$ versus $3/2$ puzzle include the following:
\begin{itemize}
\item The experimental signal for the remaining 15\% of $X_c$ is rather vague; therefore, only a small part might actually be $D_0^{1/2}$ and $D_1^{1/2}$.

\item Sum rules like (\ref{EQN001}) might not be saturated by the ground states.

\item Sum rules derived by OPE hold in the static limit and might change for finite heavy quark masses.

\item Sum rules make statements about the zero recoil situation ($w = 1$), where the $B$ and the $D$ meson have the same velocity; to obtain decay rates, however, one has to integrate over $w$.
\end{itemize}

With a dynamical lattice computation of $\tau_{1/2}(1)$ and $\tau_{3/2}(1)$ in the static limit, which is presented in the following section, we attempt to shed some light on this puzzle.


\section{Lattice computation of $\tau_{1/2}$ and $\tau_{3/2}$}

For a more detailed presentation of this computation we refer to \cite{Blossier:2009vy}. We use a method, which was proposed and tested in the quenched case in \cite{Becirevic:2004ta}.

Since the ``Isgur-Wise relations'' (\ref{EQN002}) and (\ref{EQN003}) are not directly useful to compute $\tau_{1/2}(1)$ and $\tau_{3/2}(1)$ (the right hand sides vanish at zero recoil), they have to be rewritten as shown in \cite{Leibovich:1997em}:
\begin{eqnarray}
\label{EQN004} & & \hspace{-0.7cm} \langle D_0^{1/2}(v) | \bar{c} \gamma_5 \gamma_j D_k b | B(v) \rangle \ \ = \ \ -i g_{j k} \Big(m(D_0^{1/2}) - m(B)\Big) \tau_{1/2}(1) \\
 & & \hspace{-0.7cm} \langle D_2^{3/2}(v,\epsilon) | \bar{c} \gamma_5 \gamma_j D_k b | B(v) \rangle \ \ = \ \ +i \sqrt{3} \epsilon_{j k} \Big(m(D_2^{3/2}) - m(B)\Big) \tau_{3/2}(1) .
\end{eqnarray}

We compute $\tau_{1/2}$ by means of (\ref{EQN004}) and an ``effective form factor'':
\begin{eqnarray}
 & & \hspace{-0.7cm} \tau_{1/2}(1) \ \ = \ \ \lim_{t_0-t_1 \rightarrow \infty \, , \, t_1-t_2 \rightarrow \infty} \tau_{1/2 , \textrm{\scriptsize effective}}(t_0-t_1,t_1-t_2) \\
\nonumber & & \hspace{-0.7cm} \tau_{1/2 , \textrm{\scriptsize effective}}(t_0-t_1,t_1-t_2) \ \ = \\
\label{EQN007} & & = \ \ \frac{1}{Z_\mathcal{D}} \left|\frac{N(P_-) N(S) \ \ \Big\langle \Big(\mathcal{O}^{(P_-)}(t_0)\Big)^\dagger \ (\bar{Q} \gamma_5 \gamma_3 D_3 Q)(t_1) \ \mathcal{O}^{(S)}(t_2) \Big\rangle}{\Big(m(P_-) - m(S)\Big) \ \ \Big\langle \Big(\mathcal{O}^{(P_-)}(t_0)\Big)^\dagger \mathcal{O}^{(P_-)}(t_1) \Big\rangle \ \ \Big\langle \Big(\mathcal{O}^{(S)}(t_1)\Big)^\dagger \mathcal{O}^{(S)}(t_2) \Big\rangle}\right| .
\end{eqnarray}
To this end we need static-light meson creation operators $\mathcal{O}^{(S)}$, $\mathcal{O}^{(P_-)}$ and $\mathcal{O}^{(P_+)}$, static-light meson masses $m(S)$, $m(P_-)$ and $m(P_+)$, 2-point and 3-point functions, and norms $N(S)$, $N(P_-)$ and $N(P_+)$. $Z_\mathcal{D}$ is a perturbatively computed renormalization constant, whose derivation is explained in detail in \cite{Blossier:2005vy,Blossier:2009vy}. The computation of $\tau_{3/2}$ is analogous. Explicit formulae can be found in \cite{Blossier:2009vy}.


\subsection{Simulation setup}

We use $L^3 \times T = 24^3 \times 48$ gauge configurations produced by the European Twisted Mass Collaboration (ETMC). The gauge action is tree-level Symanzik improved and the fermionic action $N_f = 2$ Wilson twisted mass at maximal twist yielding automatic $\mathcal{O}(a)$ improvement of physical quantities. The lattice spacing is $a = 0.0855 \, \textrm{fm}$. To be able to extrapolate our results to physical light quark masses, we consider three different bare quark masses $\mu_\mathrm{q}$ corresponding to ``pion masses'' $m_\mathrm{PS}$, which are listed in Table~\ref{TAB002}. For more details regarding these gauge configuration we refer to \cite{Boucaud:2007uk,Boucaud:2008xu}.

\begin{table}[htb]
\begin{center}
\begin{tabular}{|c|c|c|}
\hline
 & & \vspace{-0.50cm} \\
$\mu_\mathrm{q}$ & $m_\mathrm{PS}$ in MeV & number of gauge configurations \\
 & & \vspace{-0.50cm} \\
\hline
 & & \vspace{-0.50cm} \\
$0.0040$ & $314(2)$ & $1400$ \\
$0.0064$ & $391(1)$ & $1450$ \\
$0.0085$ & $448(1)$ & $1350$\vspace{-0.50cm} \\
 & & \\
\hline
\end{tabular}
\end{center}

\caption{\label{TAB002}Bare quark masses, pion masses and number of gauge configurations.}
\end{table}


\subsection{Static-light meson creation operators}

The meson creation operators we use are latticized versions of the continuum expression
\begin{eqnarray}
\label{EQN005} \mathcal{O}^{(\Gamma)}(\mathbf{x}) \ \ = \ \ \bar{Q}(\mathbf{x}) \int d\hat{\mathbf{n}} \, \Gamma(\hat{\mathbf{n}}) U(\mathbf{x};\mathbf{x} + r \hat{\mathbf{n}}) \psi^{(u)}(\mathbf{x} + r \hat{\mathbf{n}}) ,
\end{eqnarray}
where $\bar{Q}(\mathbf{x})$ creates a static antiquark at position $\mathbf{x}$, $\psi^{(u)}(\mathbf{x} + r \hat{\mathbf{n}})$ creates a light quark separated by a distance $r$ from the static antiquark, $U$ is a gauge covariant parallel transporter and $\Gamma$ a combination of spherical harmonics and $\gamma$ matrices yielding well defined parity $\mathcal{P}$ and total angular momentum of the light degrees of freedom $j$. The operators are collected in Table~\ref{TAB003}.

\begin{table}[htb]
\begin{center}
\begin{tabular}{|c||c|c||c|c||c|}
\hline
 & & & & & \vspace{-0.50cm} \\
$\Gamma(\hat{\mathbf{n}})$ &  $J^\mathcal{P}$ & $j^\mathcal{P}$ & $\mathrm{O}_\mathrm{h}$ & lattice $j^\mathcal{P}$ & notation \\
 & & & & & \vspace{-0.50cm} \\
\hline
 & & & & & \vspace{-0.50cm} \\
\hline
 & & & & & \vspace{-0.50cm} \\
$\gamma_5$ & $0^-$ & $(1/2)^-$ & $A_1$ & $(1/2)^- \ , \ (7/2)^- \ , \ ...$ & $S$ \\
$1$ &  $0^+$ & $(1/2)^+$ & & $(1/2)^+ \ , \ (7/2)^+ \ , \ ...$ & $P_-$ \\
 & & & & & \vspace{-0.50cm} \\
\hline
 & & & & & \vspace{-0.50cm} \\
$\gamma_1 \hat{n}_1 - \gamma_2 \hat{n}_2$ (cyclic) & $2^+$ & $(3/2)^+$ & $E$ & $(3/2)^+ \ , \ (5/2)^+ \ , \ ...$ & $P_+$ \\
$\gamma_5 (\gamma_1 \hat{n}_1 - \gamma_2 \hat{n}_2)$ (cyclic) & $2^-$ & $(3/2)^-$ & & $(3/2)^- \ , \ (5/2)^- \ , \ ...$ & $D_\pm$\vspace{-0.50cm} \\
 & & & & & \\
\hline
\end{tabular}
\end{center}

\caption{\label{TAB003}$J$: total angular momentum; $j$: total angular momentum of the light degrees of freedom; $\mathcal{P}$: parity.}
\end{table}


\subsection{2-point functions, static-light meson masses, norms of meson states}

With meson creation operators (\ref{EQN005}) at hand it is straightforward to compute the 2-point functions
\begin{eqnarray}
\label{EQN006} \mathcal{C}^{(\Gamma)}(t) \ \ = \ \ \Big\langle \Big(\mathcal{O}^{(\Gamma)}(t)\Big)^\dagger \mathcal{O}^{(\Gamma)}(0) \Big\rangle \quad , \quad \Gamma \ \ \in \ \ \{ \gamma_5 \, , \, 1 \, , \, \gamma_1 \hat{n}_1 - \gamma_2 \hat{n}_2 \} .
\end{eqnarray}

\begin{figure}[htb]
\begin{center}
\input{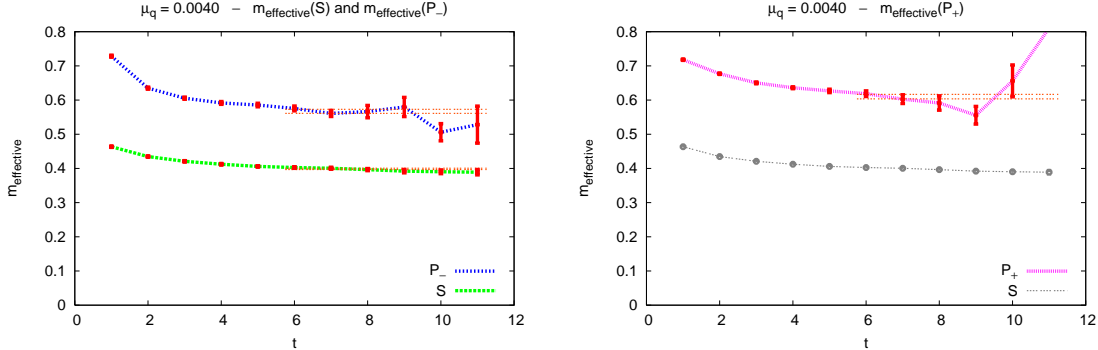}
\caption{\label{FIG001}Effective masses for $S$, $P_-$ and $P_+$ for $\mu_\mathrm{q} = 0.0040$.}
\end{center}
\end{figure}

From these 2-point functions we extract the meson masses $m(S)$, $m(P^-)$ and $m(P^+)$ via effective mass plateaus. To illustrate the quality of our data we show effective masses for $\mu_\mathrm{q} = 0.0040$ in Figure~\ref{FIG001}. For details regarding the computation of the low lying static-light meson spectrum within our twisted mass setup we refer to \cite{Jansen:2008ht,Jansen:2008si}.

Moreover, we obtain the ground state norms $N(S)$, $N(P_-)$ and $N(P_+)$ by fitting exponentials to the 2-point functions (\ref{EQN006}) at large temporal separations.


\subsection{3-point functions}

The computation of the 3-point functions is again straightforward. We chose to represent the covariant derivative inside the heavy-heavy current in a symmetric way by a single spatial link in positive and negative direction.


\subsection{Results}

\begin{figure}[htb]
\begin{center}
\input{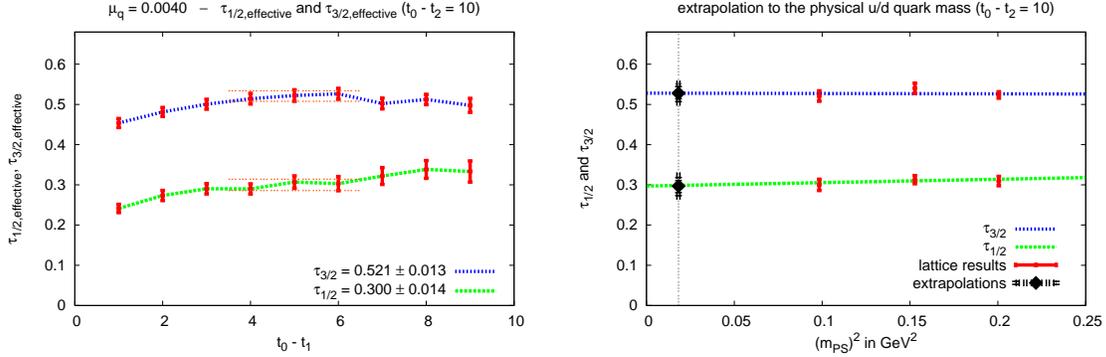}
\caption{\label{FIG002}
\textbf{a)}~Effective form factors $\tau_{1/2,\textrm{\scriptsize effective}}$ and $\tau_{3/2,\textrm{\scriptsize effective}}$ for $t_0 - t_2 = 10$ and $\mu_\mathrm{q} = 0.0040$.
\textbf{b)}~Linear extrapolation of $\tau_{1/2}$ and $\tau_{3/2}$ in $(m_\mathrm{PS})^2$ to the physical $u/d$ quark mass.
}
\end{center}
\end{figure}

In Figure~\ref{FIG002}a we show the effective form factors $\tau_{1/2,\textrm{\scriptsize effective}}$ (eqn.\ (\ref{EQN007})) and $\tau_{3/2,\textrm{\scriptsize effective}}$ for $t_0 - t_2 = 10$ as functions of $t_0 - t_1$ for $\mu_\mathrm{q} = 0.0040$ (plots for the other two quark masses look qualitatively identical). We extract $\tau_{1/2}$ and $\tau_{3/2}$ by fitting constants to the central three data points as indicated by the dashed lines. Results are collected in Table~\ref{TAB004}.

\begin{table}[htb]
\begin{center}
\begin{tabular}{|c||c|c||c|}
\hline
 & & & \vspace{-0.50cm} \\
$\mu_\mathrm{q}$ & $\tau_{1/2}(1)$ & $\tau_{3/2}(1)$ & $(\tau_{3/2})^2 - (\tau_{1/2})^2$ \\
 & & & \vspace{-0.50cm} \\
\hline
 & & & \vspace{-0.50cm} \\
$0.0040$ & $0.300(14)$ & $0.521(13)$ & $0.181(16)$ \\
$0.0064$ & $0.313(10)$ & $0.540(13)$ & $0.194(13)$ \\
$0.0085$ & $0.309(12)$ & $0.524(8) \ \, $ & $0.178(9) \ \, $\vspace{-0.50cm} \\
 & & & \\
\hline
\end{tabular}
\end{center}

\caption{\label{TAB004}$\tau_{1/2}$ and $\tau_{3/2}$ and their contribution to the Urlatsev sum rule.}
\end{table}

As expected from sum rules $\tau_{3/2}$ is significantly larger than $\tau_{1/2}$. Moreover, we find that the ground states fulfill the Uraltsev sum rule (\ref{EQN001}) by around 80\%.

We use our results at three different values of the pion mass to linearly extrapolate $\tau_{1/2}$ and $\tau_{3/2}$ in $(m_\mathrm{PS})^2$ to the physical $u/d$ quark mass ($m_\mathrm{PS} = 135 \, \textrm{MeV}$; cf.\ Figure~\ref{FIG002}b). Our final result is
\begin{eqnarray}
\label{EQN008} \tau_{1/2}^{m_\textrm{\scriptsize phys}}(1) \ \ = \ \ 0.297(26) \quad , \quad \tau_{3/2}^{m_\textrm{\scriptsize phys}}(1) \ \ = \ \ 0.528(23) .
\end{eqnarray}


\section{Conclusions}

Our result (\ref{EQN008}) confirms the sum rule expectation that $\tau_{3/2}(1) \gg \tau_{1/2}(1)$ in the static limit. When comparing to the experimentally measured form factors ($\tau_{1/2}^\textrm{\scriptsize exp}(1) = 1.28$ and $\tau_{3/2}^\textrm{\scriptsize exp}(1) = 0.75$ \cite{:2007rb}) we find fair agreement for $\tau_{3/2}$ but a strong discrepancy for $\tau_{1/2}$.

In our opinion this discrepancy calls for action both on the theoretical and the experimental side: it would be highly desirable to have a first principles lattice computation of $\tau_{1/2}$ and $\tau_{3/2}$ beyond the zero recoil situation and also for finite heavy quark masses; on the other hand a thoroughly refined experimental analysis of the decay into $1/2$ $D^{\ast \ast}$'s, for which the signal is rather faint, seems to be necessary.


\section*{Acknowledgments}

B.B.\ and O.P.\ thank Ikaros Bigi and the other authors of~\cite{Bigi:2007qp} for many discussions on these issues and having stimulated the present work. We also thank Dietmar Ebert, Vladimir Galkin, Karl Jansen, Chris Michael, David Palao, Andrea Shindler and Ruth Van de Water for many helpful discussions.

This work has been supported in part by the EU Contract No.~MRTN-CT-2006-035482, ``FLAVIAnet'', by the DFG Sonderforschungsbereich/Transregio SFB/TR9-03 and by the project ANR-NT-05-3-43577 (QCDNEXT).

We thank CCIN2P3 in Lyon and the J\"ulich Supercomputing Center (JSC) for having allocated to us computer time, which was used in this work.



\end{document}